\documentclass[twocolumn]{aastex631}

\usepackage{bm,amsmath}
\usepackage{url}
\usepackage{soul}

\begin{document}

\title{Cascaded Gamma-ray Emission Associated with the KM3NeT Ultra-High-Energy Event KM3-230213A}

\author[0000-0002-5387-8138]{Ke Fang}
\affiliation{Department of Physics, Wisconsin IceCube Particle Astrophysics Center, University of Wisconsin, Madison, WI, 53706}

\author[0000-0001-6224-2417]{Francis Halzen}
\affiliation{Department of Physics, Wisconsin IceCube Particle Astrophysics Center, University of Wisconsin, Madison, WI, 53706}

\author[0000-0001-8837-4127]{Dan Hooper}
\affiliation{Department of Physics, Wisconsin IceCube Particle Astrophysics Center, University of Wisconsin, Madison, WI, 53706}

\date{\today}

\begin{abstract}

A neutrino-like event with an energy of $\sim 220 \,{\rm PeV}$ was recently detected by the KM3NeT/ARCA telescope. 
If this neutrino comes from an astrophysical source, or from the interaction of an ultra-high-energy cosmic ray in the intergalactic medium, the ultra-high-energy gamma rays that are co-produced with the neutrinos will scatter with the extragalactic background light, producing an electromagnetic cascade and resulting in emission at GeV-to-TeV energies. In this paper, we compute the gamma-ray flux from this neutrino source considering various source distances and strengths of the intergalactic magnetic field (IGMF). We find that the associated gamma-ray emission  could be observed by existing imaging air cherenkov telescopes and air shower gamma-ray observatories, unless the strength of the IGMF is $B\gtrsim 3\times 10^{-13}$~G, or the ultra-high-energy gamma-rays are attenuated inside of the source itself. In the latter case, this source is expected to be radio-loud.  
\end{abstract}


\section{introduction}

The KM3NeT Collaboration has recently reported the detection of an approximately horizontal and extremely energetic muon track, $E_{\mu} \sim \mathcal{O}(100 \, {\rm PeV})$~\citep{KM3NeTEvent}. The event,  KM3-230213A, was observed using a configuration of 21 detection lines, which constitute about 10\% of the planned ARCA detector. The energy and orientation of this track suggest that it is not of atmospheric origin. If this event was generated by an astrophysical neutrino, it would represent the single highest-energy neutrino detected to date.

If this event was produced by a neutrino that is part of an isotropic, diffuse flux, it would be in tension with constraints that have been placed by the IceCube Neutrino Observatory~\citep{IceCube:2018fhm, IceCube:2024fxo,IceCube:2025ezc}, which has a larger effective area and has been collecting data for much longer than KM3NeT/ARCA. Such a scenario would thus require the KM3NeT event to be an unlikely upward fluctuation. 
Similarly, this event cannot be explained by a long-term, steady source. At the declination of the KM3-230213A event, and at energies of $\sim  100 \, {\rm PeV}$, the effective area of KM3NeT/ARCA will, when completed, be comparable to that of IceCube~\citep{KM3NeT:2018wnd, KM3NeT:2024uhg,Aartsen:2019fau}, yet no source has been found in this direction by IceCube after ten years of data taking~\citep{IceCube:2019cia}.

Alternatively, this event might arise from a brief, individual source of ultra-high-energy neutrinos. Other detectors could overlook such transient activity if the neutrinos arrived from a direction with large background, such as those with a high zenith angle.

A neutrino transient could occur when a short-term neutrino emitter turns on, or when a population of ultra-high-energy cosmic rays (UHECRs) is intermittently injected into the intergalactic medium, resulting in the production of cosmogentic (also referred to as Greisen–Zatsepin–Kuzmin \citealp{1966PhRvL..16..748G, 1966JETPL...4...78Z}; GZK) neutrinos. While the corresponding cosmic rays will be delayed by the IGMF for thousands of years, the neutrinos arrive without deflection or interaction. GZK neutrinos are expected to point back to the source of the UHECRs, so long as the GZK loss length is short compared to the distance to the source.

Gamma rays are co-produced with neutrinos in cosmic-ray interactions. These accompanying gamma rays could, therefore, provide a smoking gun signature for the origin of this exceptional event. In this work, we evaluate the flux of the gamma-ray counterpart of the KM3NeT/ARCA event under various assumptions about the distance to the neutrino source and the strength of the IGMF. We find that this source should be detectable by existing gamma-ray telescopes, either shortly following the neutrino event or in the coming years. We also discuss the implications of a non-detection of this gamma-ray emission, in particular, on the constraints that this would allow us place on the IGMF and on the radiation field of the source.

\section{Analytical estimations}\label{sec:analy}

The interaction length of a high-energy photon or electron in the extragalactic background light (EBL) is given by 
\begin{eqnarray}\label{eqn:interlength}
    \lambda^{-1} = \int_{-1}^{1} d\mu \, \frac{1-\mu}{2}\int_0^\infty d\epsilon \, \frac{dn}{d\epsilon} \, \sigma (E_\gamma, \epsilon),
\end{eqnarray}
where $\sigma$ is the cross section for pair production ($\gamma\gamma$) or inverse Compton (IC) scattering, for a photon or electron, respectively, $dn/d\epsilon$ is the spectrum of the target photons with energy $\epsilon$, and the outer integral averages over the interaction angles $\mu$ assuming the radiation field to be isotropic.

\begin{figure} 
    \centering
   \includegraphics[width=0.49\textwidth]{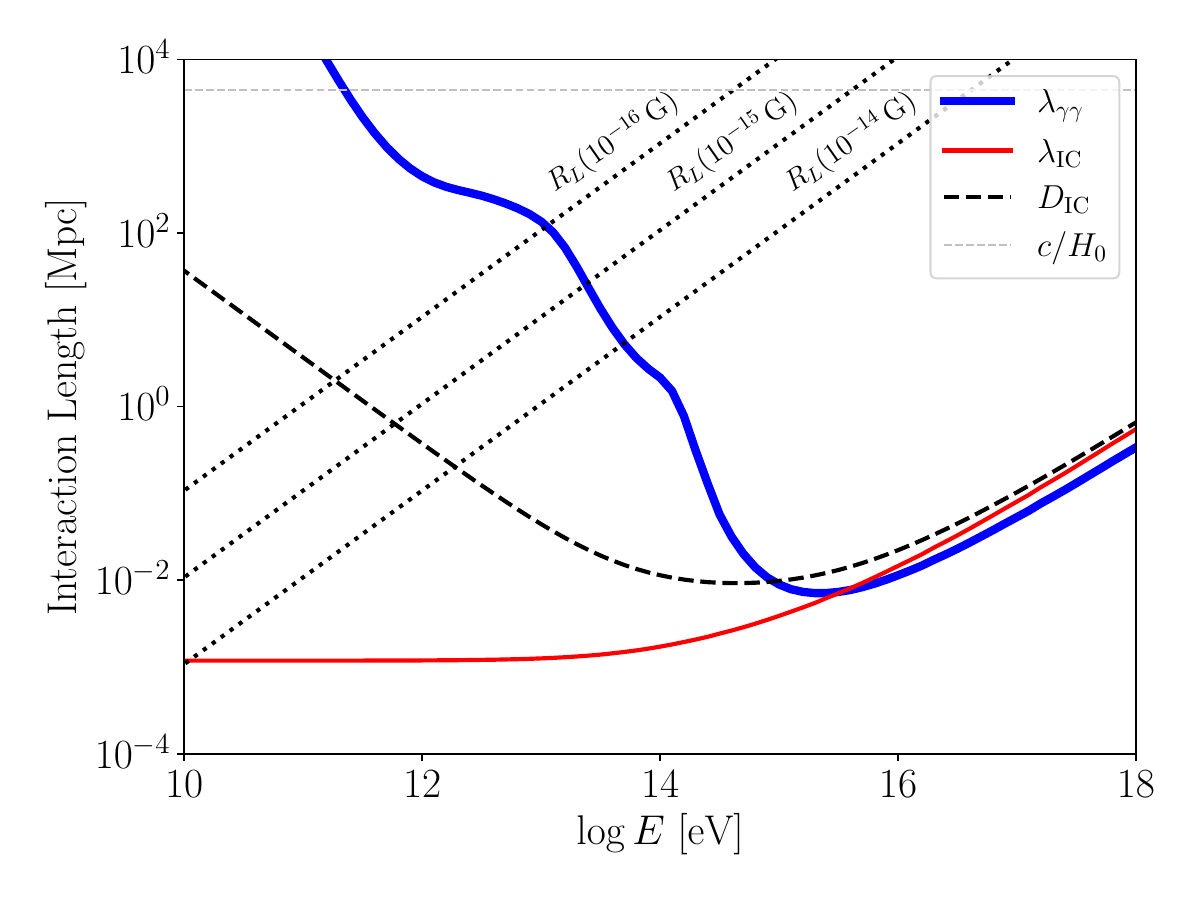}
    \caption{
    \label{fig:interactionLength} The interaction lengths of gamma rays and electrons/positrons due to pair production (blue) and inverse Compton scattering (red) in the local universe (as calculated using Eq.~\ref{eqn:interlength} with $dn/d\epsilon$ evaluated at $z=0$). For comparison, we also show the energy loss length ($D_{\rm IC}$ from Eq.~\ref{eqn:DIC};  dashed black) and the Larmor radius of electrons for three values of the strength of the intergalactic magentic field  ($R_L$;  dotted black), as well as the adiabatic energy loss length due to cosmological expansion (dotted silver).} 
\end{figure} 

In Fig.~\ref{fig:interactionLength}, we show $\lambda_{\gamma\gamma}$ and $\lambda_{\rm IC}$ as a function of the photon and electron energy. Assuming that the KM3NeT/ARCA event was generated by a neutrino with an energy of $E_\nu = 220$~PeV, the pion-decay gamma rays from the same proton interaction would have an energy of $E_{\gamma,0}  \approx 2\,E_\nu = 440 $~PeV, where the subscript denotes the initial injection of the photon at the source, prior to the formation of any cascade. 
A photon at this energy will survive only over a short distance, $\lambda_{\gamma\gamma} \approx  170$~kpc,
before undergoing pair production with the EBL. The electrons and positrons that result from this process will go on to up-scatter photons in the EBL, producing a new generation of high-energy gamma rays.
These secondary gamma rays, with slightly less energy than the primaries, enter the same interaction loop, resulting in the formation of an electromagnetic cascade. This processes ceases only when the photons reach an energy that is below the threshold for pair production, $E_{\gamma}^{\rm th} \lesssim m_e^2 / \epsilon_{\rm EBL} \approx 0.4 \, {\rm TeV} \times (\epsilon_{\rm EBL}/0.68\,{\rm eV})^{-1}$.

The electrons and positrons that are produced in a cascade are deflected by the IGMF.  In a uniform field,  or in a turbulent field with a coherence length that is much larger than the electron's cooling distance, the total deflection of an electron is given by the sum of its deflection in each displacement, $ds$: $\delta = \int_0^x ds / R_L(s)$, where $R_L = E_e /eB$ is the Larmor radius. We can write the displacement in terms of the electron energy loss rate, $ds = c (dt/dE) dE$, where $dE/dt = (4/3) c\sigma_T u_{\rm CMB} \gamma_e^2$ in the Thomson regime, and $u_{\rm CMB}$ is the energy density of the  CMB. This leads to the following analytical solution: 
\begin{eqnarray}\label{eqn:edeflection}
    \delta (E_e) &=& \int_{E_{e,0}}^{E_e} dE\, c \frac{dt}{dE} \frac{1}{R_L}  \\ \nonumber
    &=&\frac{D_{\rm IC}(E_e)}{2 R_{\rm L}(E_e)}\left[1-\left(\frac{E_e}{E_{e,0}}\right)^2\right]  \approx  \frac{D_{\rm IC}(E_e)}{2 R_{\rm L}(E_e)},
\end{eqnarray}
where $E_{e, 0}$ and $E_e$ are the electron's initial and final energies, and
\begin{equation}\label{eqn:DIC}
    D_{\rm IC}(E_e) \equiv  c  \frac{E_e}{dE/dt(E_e)}
\end{equation}
is the energy loss length of an electron at energy, $E_e$. Both $D_{\rm IC}$ and $R_L$ are shown in Fig.~\ref{fig:interactionLength}.
The approximation at the end of Eq.~\ref{eqn:edeflection} relies on the fact that electrons lose most of their energy in this propagation, $E_e  \ll E_0$.

\begin{figure} 
    \centering
   \includegraphics[width=0.49\textwidth]{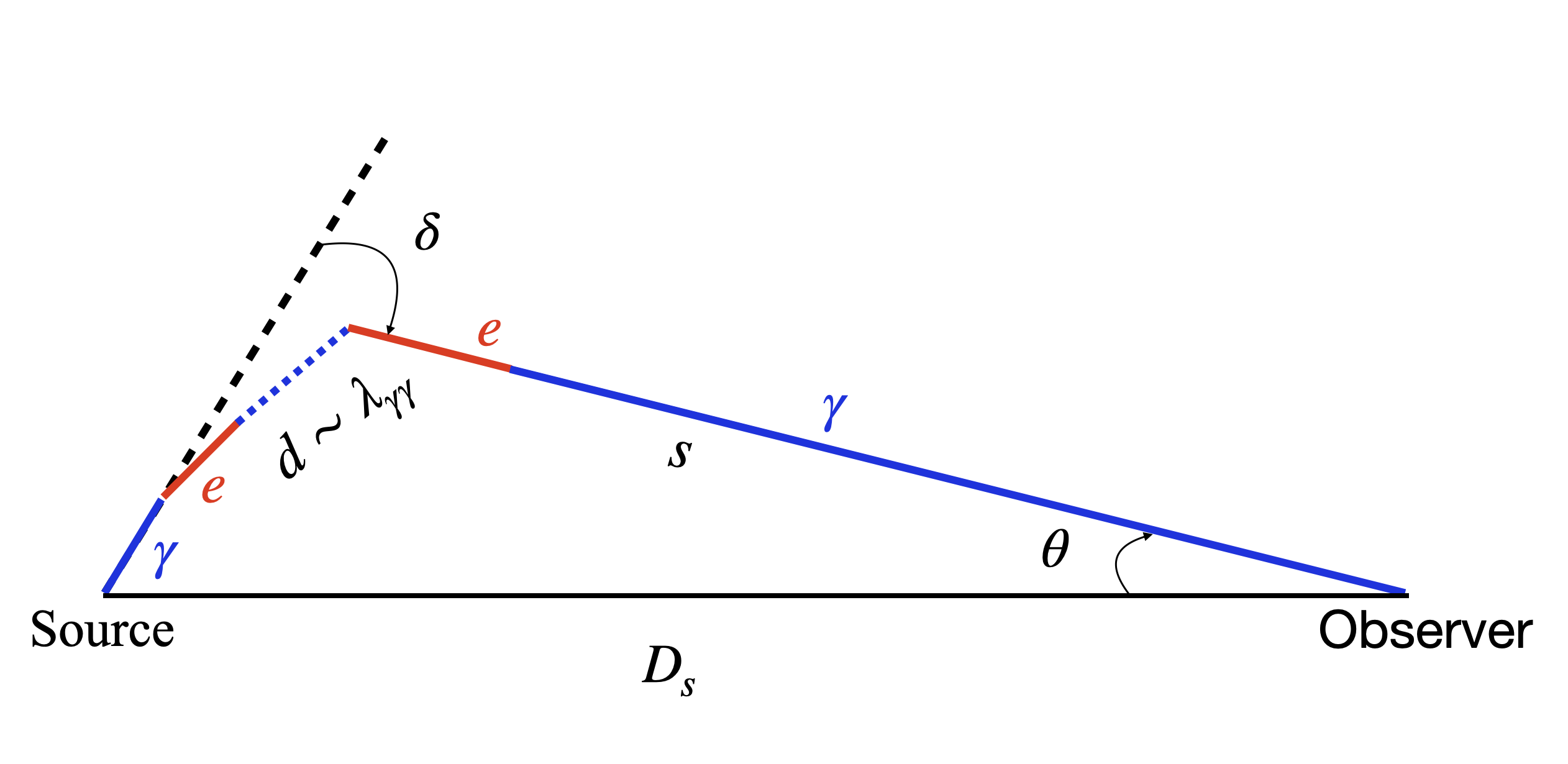}
    \caption{
    \label{fig:geometry} The geometry of the last-generation cascades.} 
\end{figure} 

Fig.~\ref{fig:geometry} describes the geometry of the last-generation cascades. A photon observed at an energy, $E_{\gamma, n}$, is  produced by an electron at energy, $E_{e,n} \equiv \gamma_{e,n}m_ec^2$, with $E_{\gamma,n} \approx (4/3) \gamma_{e,n}^2 \epsilon_{\rm CMB}$. The electron is produced at a distance, $d$, from the source. As $\lambda_{\gamma\gamma}\gg \lambda_{\rm IC}$ at $E_{\gamma, n}$, $d$ is dominated by the pair production interaction length of the previous-generation photon, $d\sim \lambda_{\gamma\gamma} (E_{\gamma, n-1})$, where $E_{\gamma, n-1} \approx 2 \,E_{e, n}$.
The deflection angle, $\theta$, of the last-generation photon follows $D_s \sin\theta = d \sin \delta$, where $D_s$ is the distance to the source. In a small-angle approximation, this simplifies to 
\begin{eqnarray}
    \theta  & \approx & \frac{\lambda_{\gamma\gamma}(E_{\gamma, n-1})}{D_s}\delta \\ \nonumber
    & \approx & 1.1'' \times \left( \frac{D_s / \lambda_{\gamma\gamma} ( E_{\gamma, n-1})}{100}\right)^{-1}\left(\frac{B}{10^{-15}\,\rm G}\right)\left(\frac{E_{\gamma, n}}{1\,\rm TeV}\right)^{-1}. 
\end{eqnarray}
Similarly, the time delay of the cascade is determined by the last-generation particles,
\begin{eqnarray}\label{eqn:dt}
    \delta t  &=& d + s - D_s\approx  \lambda_{\gamma\gamma} \frac{\delta^2}{2} \\ \nonumber
    &\approx& 4.9\, {\rm yr} \times \left(\frac{\lambda_{\gamma\gamma}(E_{\gamma, n-1})}{10\,\rm Mpc}\right)\left(\frac{B}{10^{-15}\,\rm G}\right)^2\left(\frac{E_{\gamma, n}}{ 1\,\rm TeV}\right)^{-2},
\end{eqnarray}
where $s= D_s\cos\theta - d \cos\delta$ is the distance traveled by the last-generation photons. This estimate suggests that we should expect to observe a point-like source of TeV-scale gamma rays in the direction of the KM3NeT/ARCA event on a timescale of years after the appearance of the neutrino event (for intermediate values of the IGMF strength). 
We note that this analytical estimate does not account for the spectral distribution of the inverse Compton and pair production processes. It also breaks down at low energies, where $D_{\rm IC} \gg R_L$, and does not apply to energies well beyond the pair production threshold, $E_\gamma^{\rm th}$. 

\section{Numerical Results}\label{sec:results}

In this section, we present our numerical results. We computed the evolution of the electromagnetic cascades using the ELMAG package (version~3.03; \citealp{2012CoPhC.183.1036K,Blytt:2019xad}) as well as the Monte Carlo code, Cosmological Electromagnetic Cascades simulation (CECsi; \citealp{Fitoussi:2017ble}). Our calculations use the EBL model of \citet{2011MNRAS.410.2556D}, although our results are largely insensitive to this choice. As shown in \citet{Fitoussi:2017ble}, for example, the spectral shape and intensity of the cascades are approximately universal among EBL models, although the cutoff energy can differ by up to $\sim 50\%$. 

The development of the cascades depend more sensitively on the strength and structure of the IGMF, which are largely unknown.  Here we briefly summarize the constraints on the properties of the IGMF, including those on its strength, $B$, and coherence length, $\lambda_B$. 
Direct measurement of the Faraday rotation of polarized radio emission from distant quasars requires $B\leq 10^{-11}$~G \citep{2009PhRvD..80l3012N}. Searches for the impact of primordial magnetic fields on the large-scale anisotropies of the CMB yields a similarlly stringent present-day limit, $B\lesssim 10^{-11}$~G \citep{Jedamzik:2018itu}. Finally, the non-detection of angular broadening around TeV blazars due to the deflection of charged leptons in the cascades leads to a constraint of $B\lesssim 10^{-14}-10^{-13}$ when fixing the coherence length to $\lambda_B = 1$~Mpc~\citep{VERITAS:2017gkr}. The precise value of this limit depends on the intrinsic spectra of these sources and on the EBL model that is adopted. Furthermore, these constraints do not apply to the pair halo regime, in which $B\gtrsim 10^{-12}$~G. Lower limits on $B$ have been derived from the non-detection of GeV-scale electromagnetic cascades from TeV blazars~\citep{2010Sci...328...73N, Essey:2010nd}, requiring $B\gtrsim 3\times 10^{-16}$~G. For further details, see the reviews of \citet{Durrer:2013pga} and~\citet{AlvesBatista:2021sln}. In what follows, we set $\lambda_B = 1$~Mpc and consider $3\times 10^{-16}\,{\rm G} \leq B \leq 3\times 10^{-13}\,\rm G$. We discuss the potential impact of a stronger field, or of a different coherence length, in Sec.~\ref{sec:discussion}.

To evaluate the gamma-ray flux from a short-duration source responsible for the KM3NeT/ARCA event, we first estimate the fluence of the neutrino source using the per-flavor isotropic cosmic neutrino flux reported by \citet{KM3NeTEvent}, $E_\nu^2\Phi = 5.8\times 10^{-8}\,\rm GeV\, cm^{-2}\,s^{-1}\,sr^{-1}$, 
as $E_\nu^2 \, dN/dE_\nu dA\sim E_\nu^2\Phi \, 4\pi \, \Delta t$, where $\Delta t = 335$ days is the livetime of ARCA used to obtain the isotropic flux. We assume that the neutrino is produced through photopion production, accompanied by a monochromatic gamma-ray source with a total energy release of
\begin{equation}
    E_\gamma^2 \frac{dN}{dE_\gamma} (z) = \frac{4\pi d_L^2}{1+z}\,   \, \left(4\,E_\nu^2 \frac{dN}{dE_\nu dA}\right)(z=0)\bigg\rvert_{E_\gamma = 2 (1+z) E_\nu}.
\end{equation}
The factor of $4$ accounts for the fact that neutral and charged pions are produced in approximately equal quantities in $p\gamma$ interactions and that $3/4$ of a charged pion's energy goes to neutrinos. If the neutrino instead comes from a proton-proton interaction, this factor would be $2$ instead of $4$. Although we take the injected photons to be monochromatic at 440~PeV, our results are insensitive to the spectral shape of the primary photons, as any gamma-rays with an energy above $E_\gamma^{\rm th}$ will cascade down to lower energies, thereby losing any initial spectral information. 
    
\begin{figure} [t]
    \centering
   \includegraphics[width=0.49\textwidth]{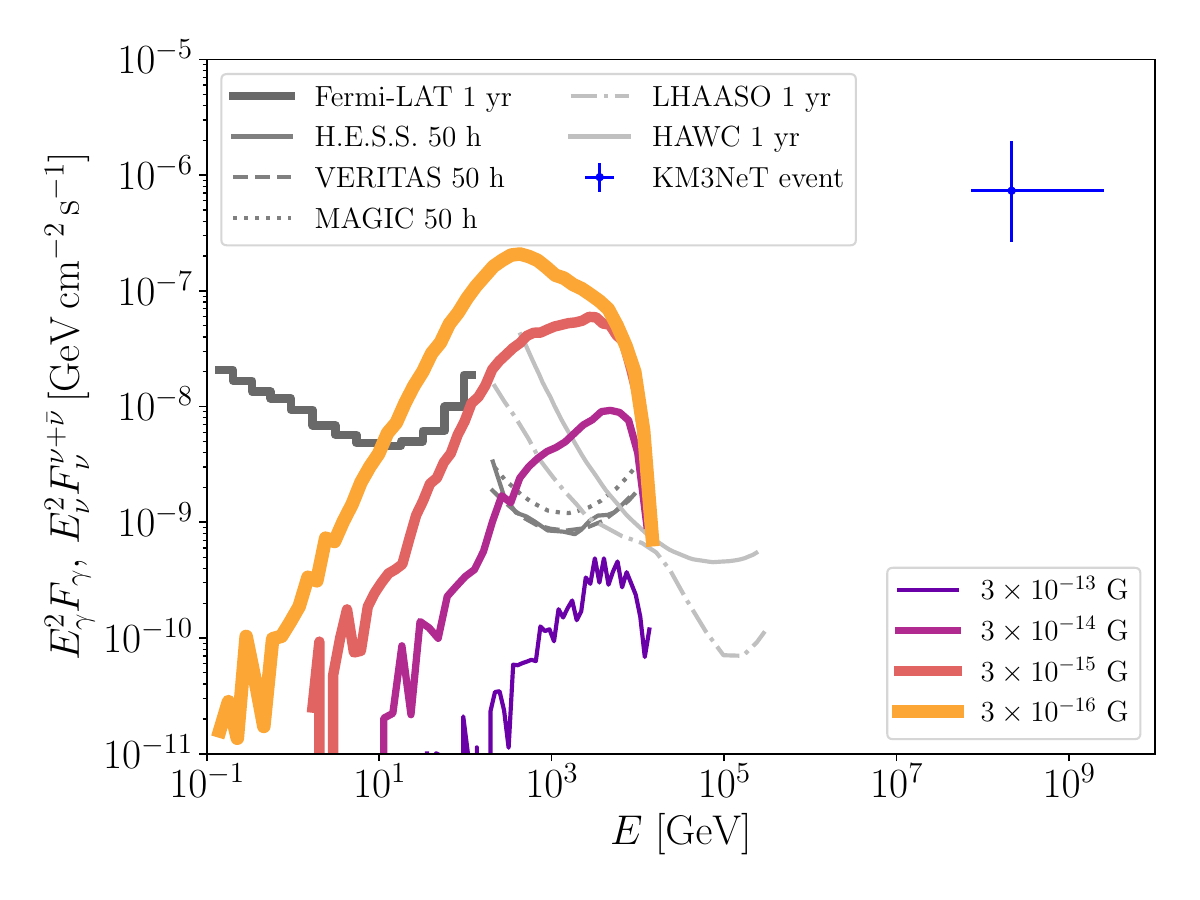}
    \caption{
    \label{fig:steady}  The flux of the gamma rays that arrive within 1 year and within $\theta = 0.1^\circ$ of the KM3NeT/ARCA event for a source at $z=0.1$, and for four values of the strength of the IGMF.  This is compared to the point-source sensitivities of {\it Fermi}-LAT (for a source at a Galactic latitude of $0^\circ$; \citealp{2012ApJS..199...31N}), the imaging atmospheric cherenkov telescopes VERITAS \citep{VERITASsen}, MAGIC \citep{Abe:2023kvd}, and H.E.S.S. \citep{CTAObservatory:2022mvt}, and the air shower gamma-ray observatories HAWC \citep{HAWC:2024plu} and LHAASO \citep{LHAASO:2019qtb}. The blue cross indicates the per-flavor neutrino flux corresponding to the KM3NeT event \citep{KM3NeTEvent} in the central 90\% neutrino energy range and $1\sigma$ flux confidence intervals. 
    } 
\end{figure}

In Fig.~\ref{fig:steady}, we show the time-integrated flux of the cascaded gamma rays in the first year following the neutrino event for a source at a distance of $z = 0.1$. We show these results for four choice of the strength of the IGMF, between $B = 3\times 10^{-16}$~G and $B = 3\times 10^{-13}$~G. 
Comparing these predicted gamma-ray fluxes to the point-source sensitivities of {\it Fermi}-LAT, imaging atmospheric cherenkov telescopes (IACTs), and air shower gamma-ray observatories, we find that the cascade from a gamma-ray emitter at this redshift is predicted to be detectable for all $B\lesssim 10^{-14}$~G, after one year of observation.  
In the case of a strong magnetic field, a longer observation time helps to better observe the event because the cascades are more delayed relative to the arrival time of the neutrino (see Eq.~\ref{eqn:dt}).

\begin{figure} [t]
    \centering
   \includegraphics[width=0.49\textwidth]{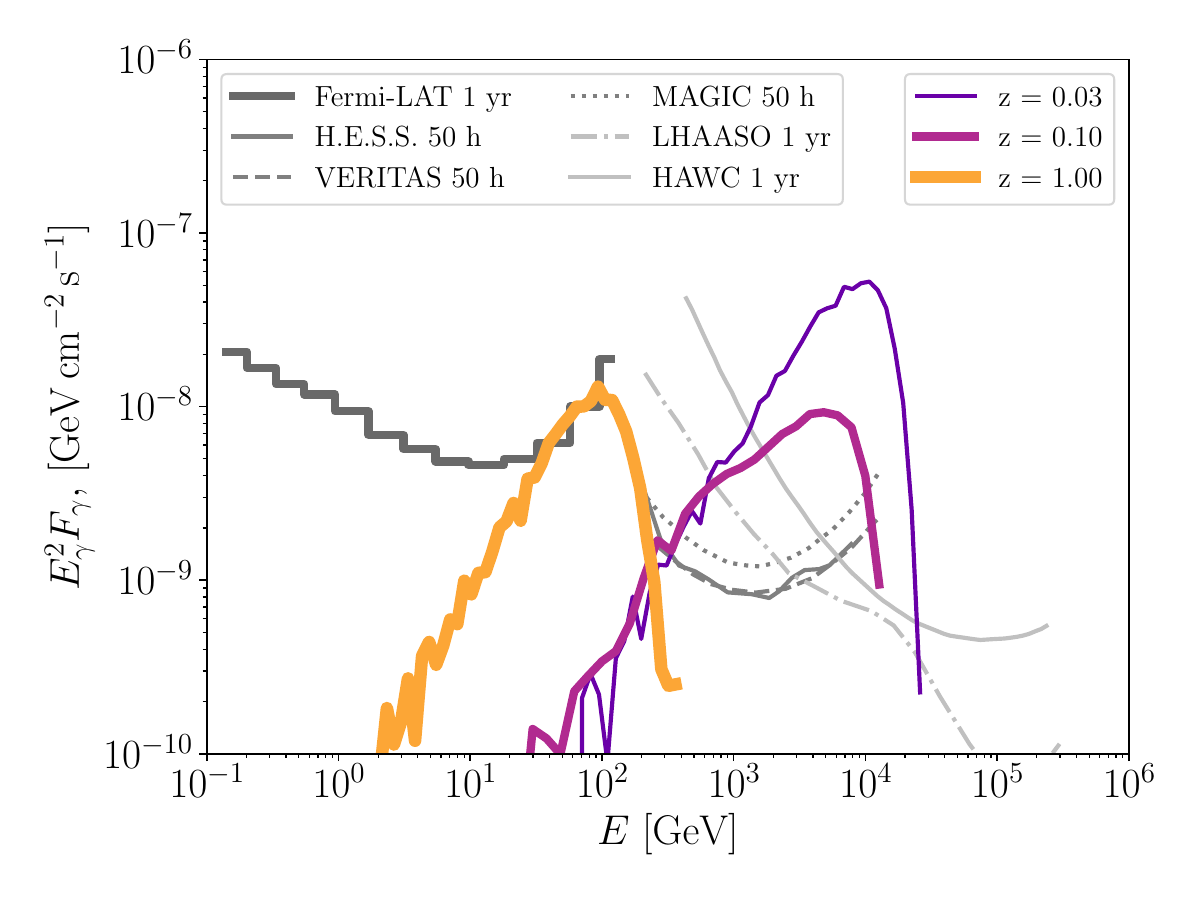}
    \caption{
    \label{fig:steadyDistance}  As in Fig.~\ref{fig:steady}, but   for three source redshifts, $z = 0.03$, 0.1, and 1, and for an IGMF strength of $B =  3\times 10^{-14}$~G.  } 
\end{figure}

Next, we consider the case of an intermediate field strength of $B = 3\times 10^{-14}$~G and evaluate the flux of the cascade  for three choices of the source redshift. As shown in Fig.~\ref{fig:steadyDistance}, for nearby sources ($z=0.03$), the resulting cascades are not fully developed, and a hard-spectrum of gamma-ray emission is predicted at energies extending up to tens of TeV (see Eq.~\ref{eqn:dt}). At intermediate redshifts ($z=0.1$), multi-TeV gamma rays are still expected, but with a somewhat softer spectrum. For more distant sources ($z = 1$), the gamma rays have fully cascaded down to the pair production threshold, $E_\gamma^{\rm th} = m_e^2 c^4  / (1+z) \epsilon_{\rm EBL}^z$, where $\epsilon_{\rm EBL}^z$ is the EBL photon energy at redshift, $z$. For $z=1$, the cascade spectrum peaks around $E_\gamma^{\rm th}\sim 100$~GeV. The spectrum of the cascade is universal and insensitive to the source distance for all $z \gtrsim 0.3$~\citep{Berezinsky:2016feh}. 
Comparing these predicted gamma-ray fluxes to the sensitivities of existing gamma-ray telescopes, we find that the cascade is expected to be observable, even if the source is located at a cosmological distance.

\begin{figure} [t!]
    \centering
   \includegraphics[width=0.49\textwidth]{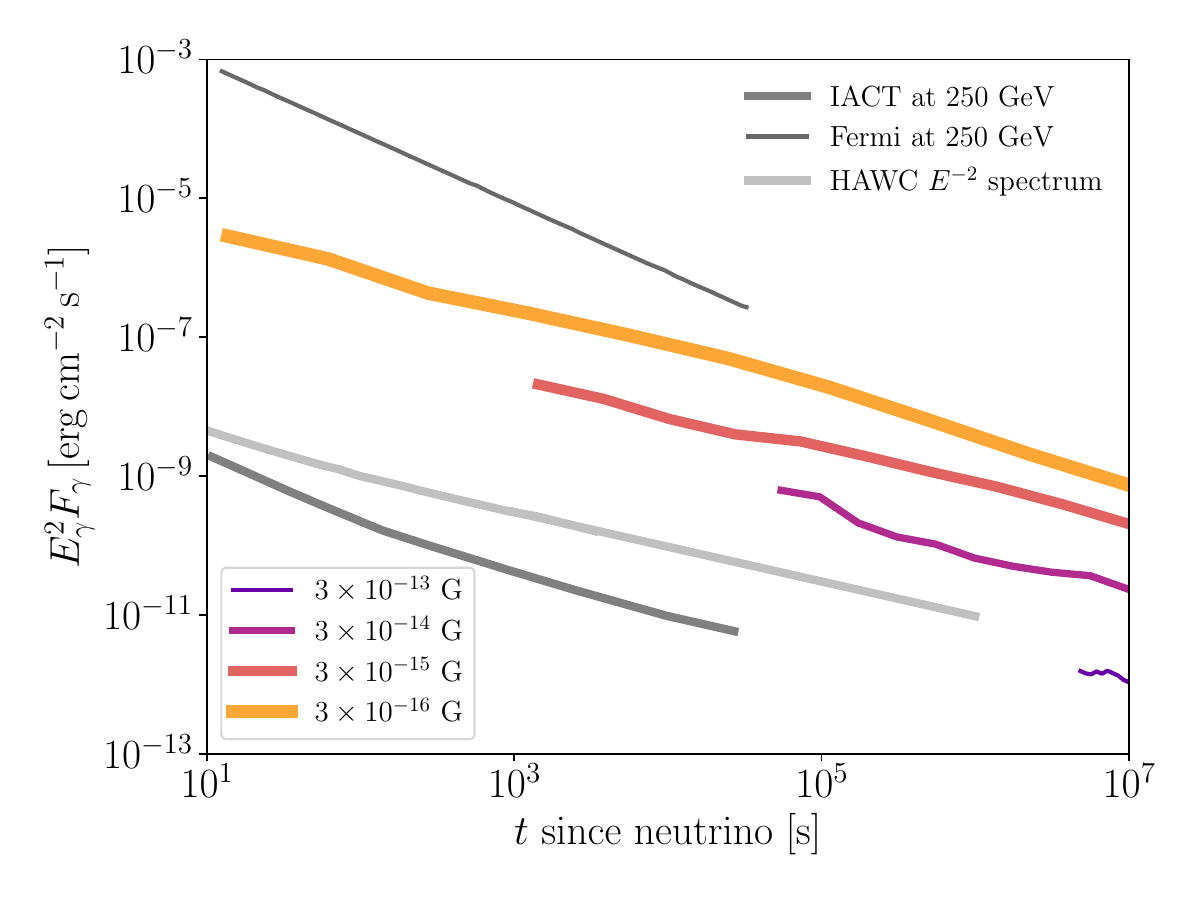}
    \caption{
    \label{fig:transientTime} The peak flux of a gamma-ray transient source associated with the KM3NeT/ARCA event, as a function of the emission duration. The source redshift is assumed to be $z=0.1$. 
    These fluxes are compared to the $5\sigma$ detection sensitivities of IACTs~(differential sensitivities of VERITAS and MAGIC at 250~GeV; \citealp{Fioretti:2019ane}), HAWC~(assuming an $E^{-2}$ spectrum; \citealp{HAWC:2011gts}), and {\it Fermi}-LAT~(at 250~GeV; \citealp{Fioretti:2019ane}) to a point-like transient source at a zenith angle of $20^\circ$.} 
\end{figure}

In Fig.~\ref{fig:transientTime}, we present the differential flux of the gamma-ray source associated with the KM3NeT/ARCA event at its peak energy, as a function of the observation time, and for various IGMF strengths. The starting time is chosen to be the point at which at least $0.01\%$ of the entire photon population has arrived. In practice, photons may arrive even earlier, although the flux evaluation would be subject to larger Poisson errors. 
Again, we compare these fluxes to the sensitivity of IACTs~\citep{Fioretti:2019ane}, HAWC~\citep{HAWC:2011gts}, and {\it Fermi}-LAT \citep{Fioretti:2019ane}. The HAWC curve beyond $t=10^3$~s is an extrapolation using $E_\gamma^2 F_\gamma \propto t^{-1/2}$. The HAWC sensitivity calculation in \citet{HAWC:2011gts} assumes a spectrum of $E^{-2}$. The IACT curves correspond to the differential sensitivities of MAGIC and VERITAS at 250~GeV. The gamma-ray flux of the neutrino transient may peak at TeV energies at early times, but the sensitivity of IACTs is not very different in that case~\citep{2015ICRC...34..981S}. 
These results suggest that a transient source responsible for the KM3NeT/ARCA event would be observable by gamma-ray telescopes hours-to-days after the neutrino event, depending on the strength of the IGMF. The source would appear later in time and be dimmer in the presence of a stronger IGMF. If $B\gtrsim 3\times 10^{-13}$~G, such a s source would likely be below the detection thresholds of existing gamma-ray telescopes.

\section{Discussion} \label{sec:discussion}

In performing the calculations presented in this study, we have adopted a coherence length of $\lambda_B = 1$~Mpc for the IGMF. For $\lambda_B \lesssim 1\,{\rm Mpc}$, the detection angle and time delay roughly scale as $\theta\propto \lambda_B^{1/2}$ and $\delta t \propto \lambda_B$, respectively \citep{Fitoussi:2017ble}. In a large coherence length regime, $\lambda_B > 1$~Mpc, the coherence length is larger than the Larmor radius and/or the energy loss length of nearly all of the charged particles in the cascade, and thus its impact on the trajectories of the pairs saturates. The prospects for detecting the gamma-ray cascade associated with the KM3NeT/ARCA event would be improved if $\lambda_B$ is lower than our fiducial value, and roughly unchanged for larger values of $\lambda_B$. 

Our main results apply to the case of $B \lesssim 3\times 10^{-13}$~G. Fields stronger than this value could delay the arrival time of the cascade by a decade or more. The detection of such a delayed source would likely require greater sensitivities, such as could be achieved by the Cherenkov Telescope Array Observatory (CTAO) or the Southern Wide-field Gamma-Ray Observatory (SWGO). On the other hand, the non-detection of any gamma-ray source associated with the KN3NeT/ARCA event could be used to place a lower limit on the strength of the IGMF in the direction of the neutrino event, approximately an order of magnitude more stringent than previous constraints~\citep{2010Sci...328...73N}. 

Given that the interaction length of gamma rays and electron-positron pairs at $\sim 400 \, {\rm PeV}$ is on the order of \mbox{$\sim 10^2 \, {\rm kpc}$}, the resulting cascades are expected to develop largely outside the immediate source environment, such as the host galaxy. If these gamma rays undergo pair production closer to the source, however, the larger magnetic fields could cause the pairs to become isotropized, resulting in the formation of a pair halo around the neutrino source  \citep{1994ApJ...423L...5A}. If the local field is stronger than $\sim 10$~nG, synchrotron losses would further reduce the flux of the resulting electromagnetic cascade. In that case, a synchrotron pair echo may be produced at sub-GeV energies \citep{Murase:2011yw}.

If the environment surrounding the source of the KM3NeT/ARCA event is opaque to very-high-energy gamma rays, the intensity of the resulting electromagnetic cascade could be suppressed.
The attenuation of gamma rays at $E_{\gamma} \sim 400 \, {\rm PeV}$ would require a radiation field that peaks in the radio band, $\epsilon = (m_e c^2)^2/E_\gamma \approx 140$~MHz, and with a luminosity and radius that satisfy the condition, $L \sigma_{\gamma\gamma}/4\pi R c \epsilon  > 1$. The spectral flux density of this radio source can be estimated as \mbox{$L / (4\pi d_L^2 \epsilon)  = 0.12 \,{\rm mJy} \times (R / 0.1 \,{\rm kpc}) \, (d_L / 6700\,{\rm Mpc})^{-2}$}, where $d_L = 6700$~Mpc is the luminosity distance of a source at redshift $z = 1$. Such a source would be detectable by low-frequency radio observatories such as LOFAR \citep{2013A&A...556A...2V}. 


In summary, if the ultra-high-energy track-like event KM3-230213A originates from an astrophysical neutrino source, then this signal should be accompanied by an electromagnetic cascade that will be detectable by existing gamma-ray telescopes. A non-detection of this gamma-ray emission could be used to place a stringent limit on the strength of the intergalactic magnetic field. Alternatively, the lack of such emission could suggest that the source in question is opaque to $\sim 400 \, {\rm PeV}$ gamma rays, requiring the source to be radio-loud.

\vspace{2em}

\begin{acknowledgments}
This work has been supported by the Office of the Vice Chancellor for Research at the University of Wisconsin--Madison with funding from the Wisconsin Alumni Research Foundation. K.F. acknowledges support from the National Science Foundation (PHY-2238916) and the Sloan Research Fellowship. This work was supported by a grant from the Simons Foundation (00001470, KF). The research of F.H was also supported in part by the U.S. National Science Foundation under grants~PHY-2209445 and OPP-2042807.
\end{acknowledgments}

\bibliography{ref}




\end{document}